\documentclass[twocolumn,prd,preprintnumbers,amsmath,amssymb,nofootinbib,superscriptaddress]{revtex4}

\usepackage{graphicx}
\graphicspath{{FIG_Tdm/}{FIG_LambdaFSTrh/}}
\usepackage{dcolumn}
\usepackage{bm}
\usepackage{epsfig}
\usepackage{epstopdf}
\usepackage{amssymb}
\usepackage{amsmath}
\usepackage{subfigure}
\usepackage{IEEEtrantools}
\usepackage{adjustbox}
\usepackage{color}
\usepackage{commath}

                              \newlength{\strikewidth}
                              \newlength{\strikelength}
\begin{document}

\title{Quasidecoupled state for dark matter in nonstandard thermal histories}
\author{Isaac Raj Waldstein}
\email{isaac14@live.unc.edu}
\affiliation{Department of Physics and Astronomy, University of North Carolina at Chapel Hill, Phillips Hall CB 3255, Chapel Hill, North Carolina 27599, USA}
\author{Adrienne L. Erickcek}
\affiliation{Department of Physics and Astronomy, University of North Carolina at Chapel Hill, Phillips Hall CB 3255, Chapel Hill, North Carolina 27599, USA}
\author{Cosmin Ilie}
\affiliation{Department of Physics and Astronomy, University of North Carolina at Chapel Hill, Phillips Hall CB 3255, Chapel Hill, North Carolina 27599, USA}
\affiliation{Department of Theoretical Physics, National Institute for Physics and Nuclear Engineering, Magurele, P.O.Box M.G.-6, Romania}


\begin{abstract}
Dark matter drops out of kinetic equilibrium with standard model particles when the momentum-transfer rate equals the expansion rate. In a radiation-dominated universe, this occurs at essentially the same time as dark matter kinetically decouples from the plasma. Dark matter may also fall out of kinetic equilibrium with standard model particles during an \mbox{early matter-dominated era (EMDE)}, which occurs when the energy content of the Universe is dominated by either a decaying oscillating scalar field or a semistable massive particle before big bang nucleosynthesis. Until now, it has been assumed that kinetic decoupling during an EMDE happens similarly to the way it does in a radiation-dominated era.  We show that this is not the case. By studying the evolution of the dark matter temperature, we establish a quasidecoupled state for dark matter in an EMDE, during which the dark matter temperature cools faster than the plasma temperature but slower than it would cool if the dark matter were fully decoupled.
The dark matter does not fully decouple until the EMDE ends and the Universe becomes radiation dominated. We also extend the criteria for quasidecoupling to other nonstandard thermal histories and consider how quasidecoupling affects the free-streaming length of dark matter.

\end{abstract}




\maketitle         
\section{Introduction}
\label{sec:Intro}

Weakly interacting massive particles (WIMPs) are prime candidates for cold dark matter. WIMPs interact with other standard model (SM) particles solely via the electroweak force, which allows them to fall out of kinetic equilibrium with the relativistic plasma as early as one second after the big bang. This departure from equilibrium occurs when the momentum-transfer rate, $\gamma$, between WIMPs and SM particles equals the Hubble expansion rate, $H$,
\begin{equation}\label{Tkd}
\gamma(T_{\mathrm{neq}}) \equiv H(T_{\mathrm{neq}}),
\end{equation}
which defines the nonequilibrium temperature \mbox{$T_{\mathrm{neq}}$ \cite{Hofmann2001}}.\footnote{{The EMDE literature typically uses Eq.~\eqref{Tkd} to define $T_{\mathrm{kd}}$ in a generic nonstandard thermal history, but we are using Eq.~\eqref{Tkd} to define a new temperature $T_\mathrm{neq}$ in order to distinguish it from $T_{\mathrm{kd}}$ defined in Eq.~\eqref{TkdLT}.}}  After dark matter (DM) kinetically decouples from the plasma, its temperature $T_{\chi}$ begins to scale as $a^{-2}$, where $a$ is the scale factor.  It has been customary to define the kinetic decoupling temperature for DM based on the late-time behavior of $T_{\chi}$ \cite{Bringmann2007, Bringmann2009},
\begin{equation}\label{TkdLT}
T_{\mathrm{kd}} \equiv {T_{\chi}}|_{T\rightarrow 0} \, \left(\frac{a}{a_\mathrm{kd}}\right)^{2},
\end{equation}
where $T$ is the plasma temperature and $a_\mathrm{kd}$ is the value of the scale factor when $T = T_\mathrm{kd}$.

In a radiation-dominated (RD) universe, Eqs.~\eqref{Tkd} and \eqref{TkdLT} imply that $T_\mathrm{kd} = T_\mathrm{neq} / K_n$, where $K_n$ is a numerical factor of order unity that depends on how the velocity-averaged DM scattering cross section scales with temperature ($\langle \sigma v \rangle\, \propto T^{n}$): \mbox{$K_2 \simeq 1.03$} for $p\,$-wave \mbox{scattering \cite{Visinelli2015, WE2017}}. Therefore, in a RD universe, the temperature at which DM falls out of equilibrium with the plasma is essentially the same as the kinetic decoupling temperature.  The timing of kinetic decoupling sets the cutoff scale in the matter power spectrum \mbox{\cite{Hofmann2001, Green2004, Green2005, Loeb2005, Bert2006, Bringmann2009, Gondolo2012, Bringmann2007}.} The small-scale cutoff, in turn, fixes the mass of the smallest protohalos that can form at high redshift \mbox{\cite{Hofmann2001, Green2004, Green2005, Loeb2005, Profumo2006, Bert2006, Bringmann2009, Gondolo2012, Bringmann2007}.}

Several investigations have explored the possibility that the Universe was not radiation dominated when DM kinetically decoupled from the plasma \cite{Gelmini2008, Arcadi2011, Kane22015, Visinelli2015, Erickcek2011, Barenboim2014, Fan2014, Erickcek2015long, Erickcek2015short}. Big bang nucleosynthesis (BBN) requires the Universe to be radiation dominated at a plasma temperature of \mbox{$T \simeq 3$ MeV} \cite{Kawasaki1999, Kawasaki2000, Hannestad2004, Ichikawa2005}, but the evolution of the Universe at higher temperatures is unknown. Both delayed inflationary reheating and the presence of gravitationally coupled scalar fields support the possibility that the Universe could have been dominated by an oscillating scalar field prior to BBN. Since scalar fields that oscillate around the minimum of a quadratic potential behave like a pressureless fluid, these scenarios include an early matter-dominated era (EMDE) prior to BBN \cite{Coughlan1983, deCarlos1984, Banks1994, Banks1995_1, Banks1995_2, Acharya2008, Acharya2009, Blinov2014}. A recent review \cite{Kane2015} concluded that EMDEs are a generic consequence of gravitationally coupled scalars in string theories. EMDEs have also been explored in the context of models where the inflaton decays to a hidden sector \cite{Tenkanen2016} and where a long-lived light mediator is responsible for interactions between DM and SM particles \cite{Zhang2015}.

If DM is produced thermally from interactions with SM particles in the plasma, then an EMDE raises the value of $T_{\mathrm{neq}}$ relative to its value in a RD era, which leads to a smaller free-streaming length \cite{Gelmini2008}. However, if DM is produced nonthermally through energy injection from a decaying scalar field, then an EMDE can lower the value of $T_{\mathrm{neq}}$ and make the transition from fully coupled to fully decoupled less sharp \cite{Arcadi2011}.  Most recently, Ref.~\cite{Visinelli2015} derived analytic expressions for how the DM temperature evolves during an EMDE and other nonstandard thermal histories.

DM kinetic decoupling in an EMDE has received a great deal of recent attention because it has been shown that an EMDE enhances the small-scale matter power spectrum and boosts the abundance of microhalos if DM stops interacting with the plasma before the onset of radiation domination \cite{Erickcek2011, Barenboim2014, Fan2014, Erickcek2015long}.\footnote{If DM remains kinetically coupled during the EMDE, then the evolution of the matter and radiation perturbations generate isocurvature perturbations that enhance the small-scale matter power spectrum \cite{Choi2015}.}  Matter perturbations that enter the horizon during radiation domination grow logarithmically with the scale factor, but they grow linearly during an EMDE.\footnote{The enhanced growth of density perturbations in an EMDE may enable primordial black holes (PBHs) to form on subhorizon scales, but PBHs only constrain EMDE scenarios if the primordial power spectrum is blue tilted \cite{Georg2016}.} The resulting enhancement in the small-scale matter power spectrum significantly increases the abundance of microhalos at high redshift ($z\gtrsim 100$) \cite{Erickcek2011, Erickcek2015long}, which can boost the DM annihilation rate by many orders of magnitude \cite{Erickcek2015long}. Such a boost to the annihilation rate can bring formerly untestable DM candidates within reach of current and future observations \cite{Erickcek2015short}. 

In this short paper, we reconsider the differential equation governing the evolution of the DM temperature, $T_{\chi}$. All prior analyses of DM kinetic decoupling in an EMDE assume that $T_{\chi}\propto a^{-2}$ when $\gamma < H$, just as it does in a RD universe. We show that this is not the case: at plasma temperatures $T < T_{\mathrm{neq}}$ in an EMDE, $T_{\chi}$ decays faster than $T$, but slower than $a^{-2}$, which implies that DM does not fully decouple in an EMDE. This \mbox{\textit{quasidecoupled}} state for DM implies that the value of $T_\mathrm{neq}$ is much greater than the value of $T_\mathrm{kd}$ in an EMDE and that $T_{\mathrm{kd}}$ is \emph{not} the temperature at which DM starts cooling as $a^{-2}$. We show that $T_\mathrm{kd}$, as defined by Eq.~\eqref{TkdLT}, does not correspond to any physical transition in the evolution of $T_{\chi}$ during an EMDE. Rather, $T_\mathrm{neq}$ marks the moment that DM drops out of equilibrium with the plasma, and the reheat temperature $T_\mathrm{RH}$ marks the end of the EMDE and the onset of full kinetic decoupling, $T_{\chi} \propto a^{-2}$. Finally, we establish the criteria for quasidecoupling in other nonstandard cosmologies.

A quasidecoupled state for DM is likely to have profound implications for our understanding of DM behavior in EMDE scenarios, including the growth of density perturbations, the free-streaming length, the abundance of microhalos, and the observational signatures of an EMDE. As a first pass at understanding these ramifications, we calculate the DM free-streaming length with quasidecoupling in an EMDE. While the DM free-streaming length is still smaller than it would be in the absence of an EMDE, it is an order of magnitude greater than it would be if DM fully decoupled during the EMDE.

We begin in Sec.~\ref{sec:2} by establishing the cosmological framework that we use throughout this paper.  In Sec.~\ref{sec:3}, we solve the evolution equation for the DM temperature in a generic cosmology in the post-equilibrium limit and show how a quasidecoupled state for DM occurs during an EMDE.  We then explore the conditions for quasidecoupling in other thermal histories, and we confirm that the general solution to the differential equation for the DM temperature agrees with the solution to the differential equation for the DM temperature in the postequilibrium limit.  We also discuss the difference between the nonequilibrium temperature $T_\mathrm{neq}$, as defined by Eq.~\eqref{Tkd}, and the kinetic decoupling temperature $T_\mathrm{kd}$, as defined by Eq.~\eqref{TkdLT}, in an EMDE. In Sec.~\ref{sec:4} we use a piecewise model for the DM velocity to study the effect of quasidecoupling in an EMDE on the free-streaming length of DM. Finally, we summarize our results in Sec.~\ref{sec:5}. The appendix presents a proof of the asymptotic series expansion for the upper incomplete gamma function that enters into the general solution for the DM temperature in a generic cosmology. We use natural units ($\hbar=c=1$) throughout this work.

\section{The Cosmological Framework}
\label{sec:2}

We employ the same cosmological framework as Ref.~\cite{Visinelli2015}: the plasma temperature and scale factor are related by $T \propto a^{-\alpha}$, where $\alpha$ is a positive constant. The Hubble rate $H$ is
\begin{equation}\label{hubblerate}
H = H_i \left(\frac{T}{T_i}\right)^{\nu}=\thinspace H_i \left(\frac{a_i}{a}\right)^{\alpha\nu} \propto \rho^{1/2},
\end{equation}
where $\rho$ is the energy density of the Universe, $\nu$ is a positive constant, and $H_i$, $T_i$, and $a_i$ are the expansion rate, plasma temperature, and scale factor, respectively, at some initial reference point. During a RD era, $\alpha =1$ and $\nu=2$ if we neglect changes in the number of relativistic degrees of freedom as the Universe cools. During an EMDE, entropy is generated by a decaying scalar field, so that the energy density of radiation scales as $\rho_{R} \propto a^{-3/2}$, which implies $\alpha = 3/8$ \cite{Giudice2001}. Since the scalar field behaves like a pressureless fluid during an EMDE, its energy density scales as $\rho_{\phi} \propto a^{-3}$, which forces $H\propto a^{-3/2}$ and implies that $\nu = 4$ in an EMDE. In our analysis, we restrict to thermally produced DM and neglect any energy transfer into DM from the decaying scalar field.

An EMDE ends when the scalar field's decay rate $\Gamma_{\phi}$ is approximately equal to the Hubble rate.  At that time, the scalar field's energy density begins to decrease exponentially, and the Universe becomes radiation dominated shortly thereafter.  It is customary to define the reheat temperature $T_\mathrm{RH}$ by 
\begin{equation}\label{Trh}
\Gamma_{\phi} \equiv \sqrt{\frac{8 \pi^{3} g_{*}(T_\mathrm{RH})}{90}} \frac{T_\mathrm{RH}^{2}}{m_\mathrm{pl}},
\end{equation}
where $m_\mathrm{pl} = G^{-1/2}$ is the Planck mass and \mbox{$g_{*}(T) = \rho_{R}(T)/[(\pi^2/30)T^4]$} is the number of relativistic degrees of freedom.  It is also convenient to define $a_\mathrm{RH}$ as the value of the scale factor when $\Gamma_{\phi} = H_i(a_i/a)^{3/2}$, 
\begin{equation}\label{arh}
\frac{a_\mathrm{RH}}{a_i} \equiv \left(\frac{H_i}{\Gamma_\phi}\right)^{2/3}.
\end{equation}
Since the transition from scalar domination to radiation domination is not instantaneous, these definitions of $T_\mathrm{RH}$ and $a_\mathrm{RH}$ do not imply that $T(a_\mathrm{RH}) = T_\mathrm{RH}$.  On the contrary, numerically solving the equations that govern the evolution of the scalar and radiation energy densities reveals that $T(a_\mathrm{RH})\simeq 0.74 T_\mathrm{RH}$.  

The kinetic decoupling of DM is governed by the elastic collision rate $\Gamma = \langle \sigma v \rangle\thinspace n_{\mathrm{rel}}$ between DM and relativistic particles with number density $n_{\mathrm{rel}}$. In most WIMP models, the $p\thinspace$-wave $(n=2)$ scattering channel dominates \cite{Bert2006}, but we maintain arbitrary $n$ for generality. Relative to the elastic collision rate, the momentum-transfer rate, $\gamma$, is suppressed by a factor of $(T/m_{\chi})$, where $m_{\chi}$ is the mass of the DM particle: $\gamma \simeq (T/m_{\chi})\, \Gamma$. The suppression factor $(T/m_{\chi})$ encodes the fact that it takes $(m_{\chi}/T) \gg 1$ elastic scatterings to appreciably alter the DM particle's momentum \cite{Hofmann2001}. Thus, the momentum-transfer rate is given by 
\begin{equation}\label{transferrate}
\gamma = \gamma_i \left(\frac{T}{T_i}\right)^{4+n}=\thinspace  \gamma_i \left(\frac{a_i}{a}\right)^{\alpha (4+n)},
\end{equation}
where $\gamma_i = \gamma(T_i)$. Lastly, since $T_{\mathrm{neq}}$ is defined by Eq.~\eqref{Tkd}, we denote the equilibrium regime by $ (\gamma/H) \gg 1$ and the postequilibrium regime by $ (\gamma/H) \ll 1$. The equilibrium regime describes the period in which \mbox{$T_{\chi}(a)\simeq T(a)$} and the postequilibrium regime includes all phases during which \mbox{$T_{\chi}(a)$} diverges from \mbox{$T(a)$}.

\section{The postequilibrium Behavior of the DM temperature}
\label{sec:3}

Using a Fokker-Planck equation for the DM particle occupation number $f_{\chi}$ with the approximation $1\pm f_{\chi} \approx 1$, it can be shown that $T_{\chi}$ satisfies the differential equation \cite{Bert2006, Bringmann2007}
\begin{equation}
a \thinspace \frac{dT_{\chi}}{da}  +  2\,T_{\chi}(a) \left[1 + \frac{\gamma(a)}{H(a)} \right] = 2 \,\frac{\gamma(a)}{H(a)}\,T(a).\label{fullode}
\end{equation}
In the postequilibrium regime, $(\gamma/H) \ll 1$, so if we drop the $(\gamma/H)$ term on the lhs of Eq.~\eqref{fullode} and assume that $\left(\gamma/H\right)\, T \ll T_{\chi}$, then $T_{\chi} \propto a^{-2}$, and DM fully kinetically decouples from the plasma. However, the assumption that  $\left(\gamma/H\right)\, T \ll T_{\chi}$ is not always valid, even in regimes where $(\gamma/H) \ll1$. To demonstrate this point, we solve Eq.~\eqref{fullode} in the postequilibrium limit
\begin{equation}\label{latetimeODE}
a \thinspace \frac{dT_{\chi}^{\mathrm{PE}}}{da}  +  2\,T_{\chi}^{\mathrm{PE}}(a) \simeq 2\,\frac{\gamma(a)}{H(a)} \thinspace T(a).
\end{equation} 
We can rewrite $T(a)$  and $(\gamma/H)$, respectively, as \mbox{$T(a) = T_{\mathrm{neq}} \thinspace \left(a_{\mathrm{neq}}/a\right)^{\alpha}$} and $ (\gamma/H) =  \left(a_{\mathrm{neq}}/a\right)^{\alpha\beta}$, where $a_{\mathrm{neq}}$ is the value of the scale factor when $T = T_\mathrm{neq}$ and $\beta \equiv (4+n - \nu)$. We restrict to the case $\beta > 0$ to guarantee that $(\gamma/H)$ monotonically decreases with time. 

If we perform the variable transformations \mbox{$g(a)\equiv (a/a_{\mathrm{neq}})^2\,T_{\chi}^{\mathrm{PE}}(a)$} and $y\equiv (a/a_{\mathrm{neq}})$, then we obtain \mbox{$(dg/dy) = 2 \, T_{\mathrm{neq}}\, y^{[1-\alpha(\beta +1)]}$}, which integrates to
\begin{equation}\label{ltsolution}
T_{\chi}^{\mathrm{PE}} \simeq \, C\, \left(\frac{a_{\mathrm{neq}}}{a} \right)^{2}\thinspace + \thinspace  \frac{2\,T_{\mathrm{neq}}}{2 - \alpha (\beta + 1)} \thinspace \left(\frac{a_{\mathrm{neq}}}{a}\right)^{\alpha(\beta +1)},
\end{equation}
where $C$ is a constant of integration. If $\alpha(\beta +1) > 2$, then Eq.~\eqref{ltsolution} shows that the $C\,(a_{\mathrm{neq}}/a)^{2}$ term dominates over the second term in the postequilibrium regime, such that \mbox{$T_{\chi}^{\mathrm{PE}} \simeq C (a_{\mathrm{neq}}/a)^{2}$}. In other words, DM fully kinetically decouples from the plasma, which also follows from Eq.~\eqref{latetimeODE} if we neglect the $(\gamma/H)\,T$ term. 

The second term in Eq.~\eqref{ltsolution} is proportional to $(\gamma/H)\,T$. If $\alpha(\beta +1) < 2$, then the second term in Eq.~\eqref{ltsolution} dominates in the postequilibrium regime, such that \mbox{$T_{\chi}^{\mathrm{PE}} \sim (\gamma/H)\,T$}, and $(\gamma/H)\,T$ in Eq.~\eqref{latetimeODE} cannot be approximated as $0$. Under these conditions, $T_{\chi}^{\mathrm{PE}}$ falls off \emph{faster} than the plasma temperature, but \emph{slower} than $a^{-2}$, which implies that DM never fully kinetically decouples from the plasma. Instead, DM enters a \mbox{\emph{quasidecoupled state}}, which represents a surprising new behavior, because all prior analyses of kinetic decoupling in nonstandard cosmologies assume that DM is fully decoupled from the plasma when $T < T_{\mathrm{neq}}$. 

\begin{table*}[t]\label{Table1}
  \centering
  \begin{tabular}{*{5}{|c|}}
  \hline
  Thermal History & $\alpha(\beta +1)$ & $\left(\gamma/H\right)\, T $  & $T_{\chi}^{\mathrm{PE}} \sim$ &  \text{Quasidecoupling?}\\
  \hline
  \hline
   \text{$\Lambda$ dom.} ($T\propto a^{-1}, H\propto T^{0}, w = -1$)  & $5+n$ &  $T_{\mathrm{neq}}\left(a_{\mathrm{neq}}/a\right)^{(5+n)}$ & $C\,(a_{\mathrm{neq}}/a)^{2}$ & No\\
   \text{Matter dom.} ($T\propto a^{-1}, H\propto T^{3/2}, w = 0$)  & $(7/2)+n$ &  $T_{\mathrm{neq}}\left(a_{\mathrm{neq}}/a\right)^{\left[(7/2)+n\right]}$ & $C\,(a_{\mathrm{neq}}/a)^{2}$ & No\\
  \text{RD} ($T\propto a^{-1}, H\propto T^{2}, w = 1/3$)  & $3+n$  & $T_{\mathrm{neq}}\left(a_{\mathrm{neq}}/a\right)^{(3+n)}$ & $C\,(a_{\mathrm{neq}}/a)^{2}$ & \text{No}\\
 \text{Kination} ($T\propto a^{-1}, H\propto T^{3}, w = 1$)  & $2+n$ &  $T_{\mathrm{neq}}\left(a_{\mathrm{neq}}/a\right)^{(2+n)}$ & $C\,(a_{\mathrm{neq}}/a)^{2}$ & No\\
 \text{EMDE}  ($T\propto a^{-3/8}, H\propto T^{4}, w = 0$)  & $(3/8) \, (1+n)$ &  $T_{\mathrm{neq}}\left(a_{\mathrm{neq}}/a\right)^{[(3/8)\,(1+n)]}$ & $T_{\mathrm{neq}}\left(a_{\mathrm{neq}}/a\right)^{[(3/8)\,(1+n)]}$ & Yes\\
  \hline
\end{tabular}
 \caption{The impact of the cosmology and scattering process on the postequilibrium behavior of DM temperature $T_{\chi}^{\mathrm{PE}}$ in various thermal histories. The equation of state parameter $w \equiv \left(P/\rho\right)$ is defined in terms of the pressure, $P$, and energy density $\rho$ of the dominant energy component. The abbreviations $\Lambda$ dom. and matter dom. denote cosmological constant-dominated and matter-dominated scenarios, respectively.} 
  \label{analysistable}
\end{table*}  

\begin{figure}
\centering
\includegraphics[width=80mm,scale=20.5]{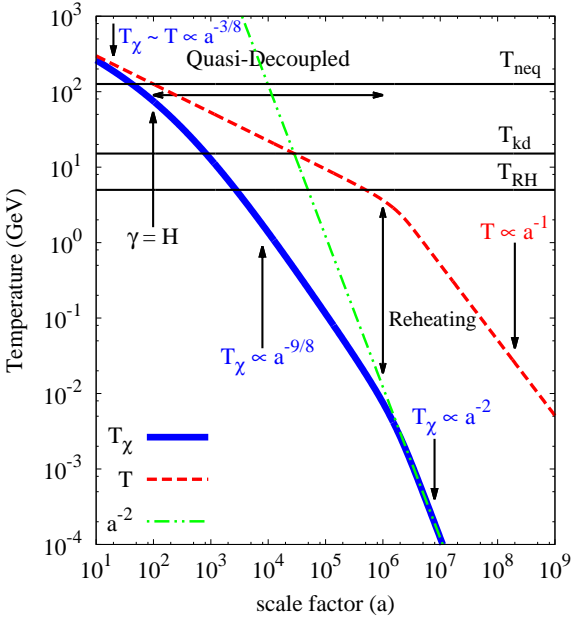}
\setlength{\abovecaptionskip}{5pt}
\setlength{\belowcaptionskip}{-10pt}
\caption{The evolution of the DM temperature $T_{\chi}$ \mbox{(solid curve)} and plasma temperature $T$ (dashed curve) vs $a$ for \mbox{$p\,$-wave} ($n = 2$) scattering in an EMDE with $T_\mathrm{RH} = 5$~GeV.  The EMDE ends and the Universe becomes radiation dominated when $a \simeq a_\mathrm{RH}$: in this figure, $a_\mathrm{RH} = 10^6$. The momentum-transfer rate $\gamma$ equals $H$ when \mbox{$T= T_\mathrm{neq} = 126.5$~GeV}, which corresponds to a DM particle that would have $T_\mathrm{neq} = 20$~GeV in a radiation-dominated universe (i.e. $T_\mathrm{kdS} = 20$~GeV). Equation~\eqref{TkdLT} gives \mbox{$T_\mathrm{kd} = 15.1$~GeV}. In the quasidecoupled regime, $T_{\chi}$ decays faster than the plasma temperature, but slower than $a^{-2}$.}\label{FIG: 1}
\end{figure}

Table \ref{analysistable} shows how the DM temperature evolves in several cosmological scenarios. An EMDE is the only cosmology listed that permits quasidecoupling. In an EMDE, \mbox{$\alpha(\beta +1) =(3/8)\,(1+n)$} is less than 2 for \mbox{$n \le 4$}. Equation \eqref{ltsolution} then demands \mbox{$T_{\chi}^{\mathrm{PE}} \propto a^{-(3/8)\,(1+n)}$}, which forces $T_{\chi}^{\mathrm{PE}}$ to decrease \textit{faster than} the plasma temperature (\mbox{$T \propto a^{-3/8}$}) but \textit{slower than} $T_\chi \propto a^{-2}$. Figure~\ref {FIG: 1} illustrates the quasidecoupled behavior of DM in an EMDE for $p\,$-wave scattering; this plot of \mbox{$T_{\chi}$} is obtained from the numerical solution to Eq.~\eqref{fullode}.  The moment when $\gamma = H$ and the onset of radiation domination (reheating) are labeled in Fig.~\ref{FIG: 1}, we see that the equilibrium regime ($\gamma \gg H$) describes the era during which \mbox{$T_{\chi}(a) \simeq T(a) \propto a^{-3/8}$}, while the postequilibrium regime includes the quasidecoupled phase, where \mbox{$T_{\chi}(a) \propto a^{-9/8}$}, and the fully decoupled phase, where \mbox{$T_{\chi}(a) \propto a^{-2}$}. 

While Eq.~\eqref{ltsolution} highlights the sensitivity of a \mbox{quasidecoupled} state to the size of \mbox{$\alpha(\beta +1)$} relative to $2$, \mbox{Table \ref{analysistable}} suggests a deeper explanation for quasidecoupling. In all thermal histories listed where entropy is conserved, such that $T \propto a^{-1}$, quasidecoupling is not permitted for \mbox{$n > 0$}. Is quasidecoupling possible in thermal histories for which entropy is conserved and is an EMDE the only entropy-producing scenario that allows quasidecoupling? To answer these questions, we recast our treatment in terms of the parameter $w \equiv P/\rho$, which we assume to be constant. First, we fix $T \propto a^{-1}$ and use $\rho \propto a^{-3(1+w)}$ to find $w = (2/3)\,\nu - 1$. Our quasidecoupling condition, \mbox{$\alpha(\beta+1) < 2$}, then implies that \mbox{$w > 1 + (2/3)\,n$}. Thus, quasidecoupling is not possible if entropy is conserved and $n\geq0$ unless $w>1$ (which does not necessarily violate causality \cite{Profumo2017, Armendariz1999, Christopherson2008}). 

In cases of entropy production, we look beyond an EMDE by searching for a condition on $w$ that permits quasidecoupling. 
If the Universe is dominated by a scalar field with energy density $\rho_{\phi}$ that decays into relativistic particles with energy density $\rho_{R}$, then
\begin{equation}\label{decayscenario}
a\,H \,\frac{d}{da}\,\rho_{R} + 4\,H\,\rho_{R} = \Gamma_{\phi}\, \rho_{\phi}.
\end{equation}
Equation \eqref{decayscenario} implies that $\rho_{R} \propto \sqrt{\rho_{\phi}}$ while \mbox{$\rho_{\phi} \gg \rho_{R}$}, which fixes $\nu = 4$ and \mbox{$\alpha = (3/8)\,(1+w)$}. Applying \mbox{$\alpha(\beta+1) < 2$} yields \mbox{$w < 16/ [3\,(1+n)]-1$}.
Therefore, any \mbox{$w<5/3$} allows quasidecoupling for \mbox{$n=1$ ($s\,$-wave)} scattering. For \mbox{$p\,$-wave scattering}, quasidecoupling is permitted for all $w < 7/9$. Table \ref{analysistable} shows that a standard kination scenario does not permit quasidecoupling, but a \emph{decaying} kination scenario (\mbox{$\nu = 4$, $w=1$}) would support quasidecoupling if the $s\,$-wave channel dominates the elastic scattering cross section.
 
We now verify that the general solution to Eq.~\eqref{fullode} is consistent with $T_{\chi}^{\mathrm{PE}}$ in Eq.~\eqref{ltsolution}. The general solution to Eq.~\eqref{fullode} with $T_{\chi}(a_i) = T(a_i)$ is \cite{Visinelli2015}
\begin{equation}\label{TCL}
T_\chi(a) = T(a)\,s^{\lambda}\,e^s\,\Gamma(1-\lambda, s),
\end{equation}
where $s(a) = [2/(\alpha \beta)](a_\mathrm{neq} /a)^{\alpha \beta}$ is the decoupling parameter for $\beta > 0$, $\lambda \equiv (2-\alpha)/(\alpha\,\beta)$, and $\Gamma(p, s)$ is the upper incomplete gamma function defined by \mbox{$\Gamma(p,s) = \int_{s}^{\infty} dt\,  t^{p-1} \, e^{-t}$} for all $p$ and all $s \ge 0$. Reference~\cite{Visinelli2015} obtained Eq.~\eqref{TCL}, but failed to notice that $T_{\chi}$ \emph{does not} scale as $a^{-2}$, i.e. that DM \emph{does not} fully decouple, when $(\gamma/H)\ll 1$ in an EMDE. However, we can coax the quasidecoupled behavior from Eq.~\eqref{TCL} by noting that $\Gamma(1-\lambda, s)$ obeys the asymptotic series expansion (see the appendix)
\begin{equation}\label{series}
\Gamma (1 - \lambda, s) = \Gamma(1 - \lambda) - \frac{s^{1-\lambda}}{1 - \lambda}\,+{\cal O}(s^{2-\lambda}),
\end{equation}
where $\Gamma(p) = \Gamma(p,0) = \int_{0}^{\infty} dt \,t^{p-1}\, e^{-t}$ for $p>0$ and $\Gamma(p) = (1/p)\,\Gamma(p+1)$ for $p<0$. Equation~\eqref{series} shows that $\Gamma(1-\lambda, s)$ approaches a constant value as $s\rightarrow 0$ if $(1-\lambda)  > 0$, which is consistent with the RD scenario for all $n > 0$. However, $\Gamma(1-\lambda, s)$ \emph{diverges} as $s\rightarrow 0$ if $(1-\lambda)  < 0$, which is equivalent to $\alpha\,(\beta+1) < 2$. Using $e^{s} \simeq 1$ for small $s$ and substituting Eq.~\eqref{series} and $s(a) = \left[2/(\alpha\beta)\right] \left(a_{\mathrm{neq}}/a\right)^{\alpha\beta}$ into Eq.~\eqref{TCL} yields
\begin{align}\label{sandlatetime}
T_\chi & \simeq 
\left[\left(\frac{2}{\alpha\beta}\right)^{\lambda} T_{\mathrm{neq}} \, \Gamma(1-\lambda)\right] \left(\frac{a_{\mathrm{neq}}}{a}\right)^{2}
\\
& + \frac{2\,T_{\mathrm{neq}}}{2 - \alpha (\beta + 1)} \left(\frac{a_{\mathrm{neq}}}{a}\right)^{\alpha(\beta +1)}\left[1+{\cal O}\left(\left[\frac{a_{\mathrm{neq}}}{a}\right]^{\alpha\beta}\right)\right]\nonumber,
\end{align}
which is consistent with Eq.~\eqref{ltsolution} for $T_{\chi}^{\mathrm{PE}}$ and fixes the constant of integration, $C$.

In a RD universe, the values of $T_\mathrm{neq}$ and $T_\mathrm{kd}$ are nearly identical. In an EMDE, the value of $T_\mathrm{neq}$ is much greater than the value of $T_\mathrm{kd}$, as demonstrated in Fig.~1. In an EMDE, Eq.~\eqref{TkdLT} implies that \mbox{$T_\mathrm{neq} > T_\mathrm{kd} > T_\mathrm{RH}$}. Moreover, $T_\mathrm{kd}$ does not mark a transition in the evolution of $T_{\chi}$. For small values of $\gamma$, corresponding to small scattering cross sections, the quasidecoupled stage is very long ($T_\mathrm{neq}\gg T_\mathrm{RH}$), so treating the DM as fully coupled and then fully decoupled with a transition point marked by $T_\mathrm{kd}$ is inappropriate. 

We can use the numerical solutions to Eq.~\eqref{fullode} to determine the discrepancy between $T_\mathrm{neq}$ and $T_\mathrm{kd}$ in an EMDE. These numerical solutions reveal that \mbox{$T_{\chi}|_{T\rightarrow 0} = \kappa_{1}\, T_{\chi}(a_\mathrm{RH})\, (a/a_\mathrm{RH})^{-2}$}, where \mbox{$\kappa_{1} = 1.37$}, $a_\mathrm{RH}$ is defined by Eq.~\eqref{arh}, and $T_\chi(a)$ is given by Eq.~\eqref{TCL}.\footnote{Equation~\eqref{TCL} and the following expressions in this section assume that the number of relativistic degrees of freedom, $g_{*}(T)$, is constant. If $g_*(T)$ changes, then Eq.~\eqref{TCL} is no longer an exact solution to Eq.~\eqref{fullode}.  However, using \mbox{$s=[2/(\alpha\beta)](T/T_\mathrm{neq})^{4+n}(a/a_\mathrm{neq})^{3/2}$} in Eq.~\eqref{TCL} provides an accurate approximate solution, and a value of $\kappa_1$ can be found for each value of $T_\mathrm{RH}$.}
During an EMDE, \mbox{$a / a_\mathrm{RH} = (2/5)^{2/3} (T_\mathrm{RH} / T)^{8/3}$} \cite{Erickcek2015long}, which implies that $T_\mathrm{kd}  = (2/5)^{4/13}\,T_\mathrm{RH} \left[T_\mathrm{RH}/[\kappa_{1}\,T_{\chi}(a_\mathrm{RH})] \right]^{3/13}$, where $T_\mathrm{RH}$ is defined by Eq.~\eqref{Trh}. Since $T_\chi$ decreases faster than $T$ while the DM is quasidecoupled, $\kappa_{1}\,T_{\chi}(a_\mathrm{RH}) \ll T_\mathrm{RH}$, and therefore $T_\mathrm{kd} \gg T_\mathrm{RH}$.

If $T_\mathrm{neq} / T_\mathrm{RH} \gg 1$, then we can approximate $T_{\chi}(a_\mathrm{RH})$ using Eq.~\eqref{sandlatetime}.  This approximation allows us to relate the value of $T_\mathrm{kd}$ to the value of $T_\mathrm{neq}$ in a RD universe, which we refer to as $T_\mathrm{kdS}$ to maintain consistency with prior investigations of kinetic decoupling in nonstandard thermal histories.  If DM decouples during an EMDE, then \mbox{$T_\mathrm{neq} = \sqrt{5/2}\, (T_\mathrm{kdS}^{2}/T_\mathrm{RH})$} \cite{Gelmini2008, Erickcek2015long}.  During an EMDE, the second term on the rhs of Eq.~\eqref{sandlatetime} is dominant for $a\gg a_\mathrm{neq}$. It follows that 
\begin{equation}
\frac{T_\mathrm{kd}}{T_\mathrm{neq}} = 0.53\, \left(\frac{T_\mathrm{RH}}{T_\mathrm{kdS}}\right)^{14/13}.
\end{equation}
For \mbox{($T_\mathrm{RH} / T_\mathrm{kdS}) \ll 1$}, $(T_\mathrm{kd}/T_\mathrm{neq}) \ll 1$, which confirms the generality of the temperature hierarchy \mbox{$T_\mathrm{neq} > T_\mathrm{kd} > T_\mathrm{RH}$} shown in Fig.~1. The temperatures $T_\mathrm{neq}$ and $T_\mathrm{kd}$ are vastly different in an EMDE with a sufficiently long quasidecoupled phase.

\section{Impact on the Matter Power Spectrum}
\label{sec:4}

The existence of a quasidecoupled phase could affect the evolution of DM perturbations in two ways. First, the DM perturbations may remain at least partially coupled to the perturbations in the relativistic plasma during the EMDE. The radiation density perturbations grow during the EMDE because they are sourced by an increasingly inhomogeneous scalar field, but they do not grow as quickly as the uncoupled DM perturbations \mbox{\cite{Erickcek2011, Barenboim2014, Fan2014, Choi2015}}, so any residual interactions between the plasma and the DM particles may suppress the growth of matter perturbations during the EMDE. Second, the fact that the DM temperature does not scale as $a^{-2}$ during the EMDE implies that the velocities of the DM particles, $v_{\chi}$, do not scale as $a^{-1}$, which profoundly affects the calculation of the comoving DM free-streaming horizon: \mbox{$\lambda_\mathrm{fs} = \int_{t_\mathrm{neq}}^{t_0} (v_{\chi}/a) dt$, where $t_{0}$} is the present age of the Universe. The evolution of the DM and plasma perturbations lies beyond the scope of this article, but we can estimate the impact of the quasidecoupled phase on $\lambda_\mathrm{fs}$ by assuming that \mbox{$v_\chi \propto \sqrt{T_\chi}$.}  We restrict our analysis to \mbox{$p\,$}-wave scattering in an EMDE. In this case, $T_\chi \propto a^{-9/8}$ during the quasidecoupled phase.

To isolate the effect of the EMDE on the postreheating velocities of the DM particles, we consider the velocity at the reheat temperature $T_{\mathrm{RH}}$. We also use the fact that $T_{\mathrm{neq}} \simeq T_{\mathrm{kdS}}^{2}/T_{\mathrm{RH}}$ for \mbox{$T_\mathrm{kdS}>T_\mathrm{RH}$} \cite{Gelmini2008, Erickcek2015long}.  Since $T\propto a^{-3/8}$ during the EMDE, \mbox{$v_\chi (T_\mathrm{RH}) \simeq \sqrt{(T_\mathrm{kdS}/m_\chi)}(T_\mathrm{RH}/T_\mathrm{kdS})^{5/2}$}. Therefore, the velocities of the DM particles are suppressed by a factor of $(T_\mathrm{RH}/T_\mathrm{kdS})^{3/2}$ if they  quasidecouple during an EMDE, as opposed to if they fully decouple during radiation domination. While this suppression is not as large as it would have been if the DM fully decoupled during an EMDE [in which case the suppression factor is \mbox{$(T_\mathrm{RH}/T_\mathrm{kdS})^{23/6}$~\cite{Erickcek2015long}}], it does imply that an EMDE still reduces the DM free-streaming horizon even though the DM particles do not fully decouple from the plasma. 

\begin{figure}
\centering
\includegraphics[height=60mm]{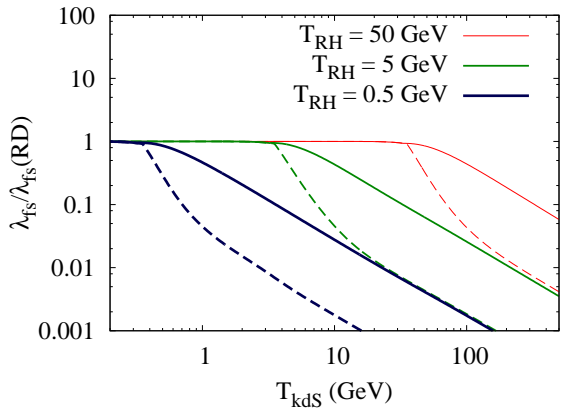}
\setlength{\abovecaptionskip}{-15pt}
\setlength{\belowcaptionskip}{-10pt}
\caption{The comoving free-streaming horizon in an EMDE divided by the comoving free-streaming horizon in a radiation-dominated universe for DM particles that would kinetically decouple at a temperature $T_\mathrm{kdS}$ in a radiation-dominated universe. If $T_\mathrm{kdS}>T_\mathrm{RH}$, then the DM kinetically decouples during an EMDE. The solid curves show how the free streaming is reduced relative to its value in a radiation-dominated universe given that the DM is quasidecoupled during the EMDE ($v_\chi \propto \sqrt{T_\chi} \propto a^{-9/16}$). The dashed curves show how the free-streaming horizon would be even smaller if the DM fully decoupled during the EMDE, as assumed in earlier investigations of EMDE scenarios \cite{Gelmini2008, Erickcek2011, Barenboim2014, Fan2014, Kane22015, Erickcek2015long, Arcadi2011, Erickcek2015short}.}
\label{FIG:FS}
\end{figure}
To quantify the impact of quasidecoupling on the free-streaming horizon, we compute 
\begin{equation}\label{fs}
\lambda_\mathrm{fs} = \int_{a_\mathrm{neq}}^{a_0} da\, \frac{v_{\chi}(a)}{a^2 H(a)}
\end{equation}
using a piecewise model for the DM velocity: \mbox{$v_\chi \simeq \sqrt{T_\mathrm{kd}/m_\chi}(a_\mathrm{kd}/a)$} for $a > a_\mathrm{RH}$ and \mbox{$v_\chi \simeq \sqrt{0.8\,T_\chi(a_\mathrm{RH})/m_\chi}(a_\mathrm{RH}/a)^{9/16}$} for $a < a_\mathrm{RH}$, where $T_\chi(a_\mathrm{RH})$ is evaluated using Eq.~(\ref{TCL}).  The factor of 0.8 in the latter expression accounts for the fact that the DM temperature obtained by numerically solving Eq.~\eqref{fullode} is $0.8$ times the value given by Eq.~\eqref{TCL} when $a = a_\mathrm{RH}$.  This model overestimates \mbox{$v_\chi$ for $a\simeq a_\mathrm{neq}$}; to compensate, we start the integration in Eq.~\eqref{fs} when \mbox{$0.8\,T_\chi(a_\mathrm{RH})(a_\mathrm{RH}/a)^{9/8} = T(a)$}, instead of at $a = a_\mathrm{neq}$. 
In contrast, if we had assumed that $T_\chi \propto a^{-2}$ while $\gamma \lesssim H$, then \mbox{$v_\chi \simeq \sqrt{(T_\mathrm{neq}/m_\chi)}(a_\mathrm{neq}/a)$} for all $T<T_\mathrm{neq}$.  \mbox{Figure 2} shows how these two models predict different free-streaming horizons if DM decouples during an EMDE.  In this figure, we plot the ratio $\lambda_\mathrm{fs}/\lambda_\mathrm{fs}$(RD), where $\lambda_\mathrm{fs}$(RD) is computed assuming that the Universe is radiation dominated when DM decouples \mbox{(at $T=T_\mathrm{kdS}$ and $a = a_\mathrm{kdS}$)} and that \mbox{$v_\chi \simeq \sqrt{(T_\mathrm{kdS}/m_\chi)}(a_\mathrm{kdS}/a)$} for all $T<T_\mathrm{kdS}$. 

Figure \ref{FIG:FS} confirms the expectations stated above: it shows that including the effects of the quasidecoupled period increases $\lambda_\mathrm{fs}$ by about an order of magnitude compared to what it would be if DM fully decoupled during an EMDE, but $\lambda_\mathrm{fs}$ is still smaller than it would have been in the absence of an EMDE.  Since the mass enclosed within the free-streaming horizon sets a lower bound on the mass of the smallest DM halos \mbox{\cite{Hofmann2001, Green2004, Green2005, Loeb2005, Profumo2006, Bert2006, Gondolo2012, Bringmann2007, Bringmann2009}}, an EMDE will still decrease the minimum halo mass if $T_\mathrm{kdS}>T_\mathrm{RH}$, but the minimum halo mass will be $\sim\!\!\,1000$ times greater than stated in Ref.~\cite{Gelmini2008}, which lays the foundations for subsequent treatments of DM free streaming in EMDE cosmologies \cite{Erickcek2011, Barenboim2014, Fan2014, Kane22015, Erickcek2015long, Arcadi2011, Erickcek2015short}.

\section{Concluding Remarks}
\label{sec:5}

We have established the existence of a quasidecoupled state for DM in thermal histories in which entropy is produced by a decaying scalar field with \mbox{$w < 16 / [3(1+n)]-1$}, where $n$ determines how the DM scattering cross section depends on temperature \mbox{($\langle\sigma v\rangle \propto T^n$)}. The nonequilibrium temperature $T_\mathrm{neq}$ is still higher than it is in a radiation-dominated era, but it marks a transition from tightly coupled to quasidecoupled as opposed to a shift from tightly coupled to fully decoupled. The DM remains quasidecoupled until the onset of radiation domination; the transition from quasidecoupled to fully decoupled does not depend on the interactions between DM and the SM. Therefore, the quasidecoupled phase should be considered as a fundamentally new state of DM.

These surprising results force us to reconsider the body of work on kinetic decoupling of DM in nonstandard thermal histories, which, until now, has been built on the assumption that DM fully decouples from the plasma when the momentum-transfer rate falls below the expansion rate. Quasidecoupling calls all previous work on the behavior of DM during an EMDE into question and merits new investigations. We have demonstrated that quasidecoupling increases $\lambda_\mathrm{fs}$ by an order of magnitude compared to the fully decoupled calculation, which has profound implications for the small-scale cutoff in the matter power spectrum, the size of the first microhalos, and the timing of their formation \cite{Erickcek2015long}. In future work, we will explore these ramifications in more detail \cite{WIE2016}.

\section*{ACKNOWLEDGMENTS}
\label{sec: Ack}
The authors thank Dan Hooper for comments on this manuscript.  This work was supported by NSF Grant No. PHY-1417446.  I.R.W. also acknowledges support from the Bahnson Fund at the University of North Carolina at Chapel Hill.

\appendix
\setcounter{secnumdepth}{0}
\section{APPENDIX: DERIVATION OF EQ.~\bf{\eqref{series}}}
\label{sec:Appendix}

The upper incomplete gamma function is given by $\Gamma(p, s) = \int_{s}^{\infty} dt \,t^{p-1}\,e^{-t}$ for all $p$ and all $s\ge0$. In this appendix, we derive Eq.~\eqref{series} for the series expansion of $\Gamma(p,s)$ for $s \ll 1$. If $p>0$, then the complete gamma function, $\Gamma(p)$, is given by 
\begin{equation*}
\Gamma(p) = \Gamma(p,0) = \int_{0}^{\infty} dt \,t^{p-1}\, e^{-t},\tag{A.1}\label{Euler}
\end{equation*}
which gives
\begin{align*}\label{seriesapp}
\Gamma(p,s) & = \left[ \int_{0}^{\infty} dt\thinspace  e^{-t} t^{p-1} -  \int_{0}^{s} dt\thinspace  e^{-t} t^{p-1}\right],\\
  & =  \Gamma(p)  - \sum_{n=0}^{\infty} \frac{(-1)^n s^{p+n}}{n!\,(p+n)},\tag{A.2}
\end{align*} 
where we used the series expansion for the $e^{-t}$ in the second integral. The series in Eq.~\eqref{seriesapp} converges for all $s$. 

If $p<0$, then $\Gamma(p)$ is defined by the recursion relation
\begin{equation*}
\Gamma(p) = \frac{1}{p}\,\Gamma(p+1).\tag{A.3}\label{rr}
\end{equation*}
We now show that it is still possible to apply Eq.~\eqref{seriesapp} for $\Gamma(p,s)$ if $p<0$ and noninteger, provided that $\Gamma(p)$ is defined by Eq.~\eqref{rr}. We follow the method described in Ref.~\cite{Bender1978} and use tautologies to decompose $\Gamma(p,s)$ as follows: 
\begin{align*}
\Gamma(p,s) & = \int_{s}^{\infty} dt\,t^{p-1} \left[\mathcal{S} + e^{-t} - \mathcal{S}\right],\\
& = \int_{s}^{\infty} dt\,t^{p-1} \mathcal{S} - \int_{0}^{s} dt\,f(t) + \int_{0}^{\infty} dt\,f(t),\\
& = I_{1} - I_{2} + I_{3}\tag{A.4}\label{decompose},
\end{align*}
where 
\begin{align*}
\mathcal{S} & \equiv \sum_{n=0}^{N} \frac{(-1)^n t^n}{n!},\\
f(t) & \equiv t^{p-1} [e^{-t} - \mathcal{S}],\\
I_{1} & \equiv\int_{s}^{\infty} dt \thinspace t^{p-1} \sum_{n=0}^{N} \frac{(-1)^n t^n}{n!},\\
I_{2} & \equiv \int_{0}^{s} dt \thinspace t^{p-1} \left[e^{-t} - \sum_{n=0}^{N}\frac{(-1)^n t^n}{n!}\right],\\ 
I_{3}& \equiv \int_{0}^{\infty} dt \thinspace t^{p-1} \left[e^{-t} -  \sum_{n=0}^{N}\frac{(-1)^n t^n}{n!}\right],
\end{align*}
and where $N$ is the largest integer less than $\abs{p}$. Integrals $I_{1}$ and $I_{2}$ can be evaluated term by term. Integral $I_{1}$ becomes
\begin{equation*}
I_{1} = \sum_{n=0}^{N}  \frac{(-1)^n}{n!} \int_{s}^{\infty} dt\thinspace t^{p+n-1} =  - \sum_{n=0}^{N} \frac{(-1)^n s^{p+n}}{n!\,(p+n)}\tag{A.5}\label{I1result}.
\end{equation*}
The integral in Eq.~\eqref{I1result} converges because \mbox{$-1 < (p+N) < 0$} demands $ (p+n) < 0$ for \mbox{$0\le n \le N$}. Using the series expansion for $e^{-t}$ and then combining the sums in $I_{2}$, integral $I_{2}$ can be written as
\begin{equation*}
I_{2}  = \sum_{n=N+1}^{\infty}\frac{(-1)^n}{n!} \int_{0}^{s} dt \thinspace t^{p+n-1} =  \sum_{n=N+1}^{\infty}   \frac{(-1)^n s^{p+n}}{n!\,(p+n)}\tag{A.6}\label{I2result}.
\end{equation*} 
The integral in Eq.~\eqref{I2result} converges because \mbox{$(p+n) > 0$} for \mbox{$N+1\le n \le \infty$}.   

Substituting the results of Eqs.~\eqref{I1result} and \eqref{I2result} into Eq.~\eqref{decompose} yields
\begin{equation*}
\Gamma(p,s) = I_{3} -  \sum_{n=0}^{\infty} \frac{(-1)^n s^{p+n}}{n!\,(p+n)}.\tag{A.7}\label{I123}
\end{equation*}
We show that $I_{3} = \Gamma(p)$ by integrating $I_{3}$ repeatedly by parts. It is simplest to start with the case of $N=1$, which corresponds to $p\,$-wave scattering in an EMDE, where \mbox{$p = -7/6$}. Suppose we integrate $I_{3}$ by parts once; then
\begin{equation*}
I_{3} = B\bigg|_{t=0}^{t=\infty} + \mathcal{I}_{3},
\end{equation*}
where the boundary term 
\begin{align*}
B  & = \frac{t^{-7/6}}{(-7/6)} \left[e^{-t} - (1-t)\right],\tag{A.8}\label{vanishatinfinity}\\
& = \frac{1}{(-7/6)} \sum_{n=2}^{\infty}\frac{(-1)^n t^{-7/6+n}}{n!},\tag{A.9}\label{boundaryterm}
\end{align*}
and 
\begin{align*}
\mathcal{I}_{3} & = \frac{1}{(-7/6)}\int_{0}^{\infty} dt \left(e^{-t} -1\right) t^{-7/6}.\tag{A.10}\label{integralcal}
\end{align*}
The $(1-t)$ term in Eq.~\eqref{vanishatinfinity} removes the first two terms from the series expansion of $e^{-t}$, so we combined it with the series expansion of $e^{-t}$ to produce Eq.~\eqref{boundaryterm}. Equation~\eqref{vanishatinfinity} makes it easy to see that $B$ vanishes at the upper limit $t=\infty$: the polynomial terms in Eq.~\eqref{vanishatinfinity} vanish at $t=\infty$ because they are both proportional to a negative power of $t$. Equation~\eqref{boundaryterm} shows that $B$ also vanishes at $t=0$ because the sum in Eq.~\eqref{boundaryterm} starts at $n=2$, such that the lowest power of $t$ is $(-7/6+2)$, which is positive. Thus $B\big|_{t=0}^{t=\infty} = 0$, which implies that $I_{3} = \mathcal{I}_{3}$.

Equation~\eqref{integralcal} for $\mathcal{I}_{3}$ is easily evaluated by integrating by parts once more: the boundary term vanishes at $t=0$ and $t=\infty$ leaving a convergent integral
\begin{align*}
I_{3} & = \frac{1}{(-7/6) (-7/6 + 1)}\int_{0}^{\infty} dt \,t^{(-7/6+2)-1} e^{-t},\tag{A.11}\label{integral}\\
& = \frac{\Gamma(-7/6+2)}{(-7/6) (-7/6+1)} = \Gamma\left(-7/6\right).\tag{A.12}\label{completegamma}
\end{align*}
Since $(-7/6+2) > 0$, the integral in Eq.~\eqref{integral} is equal to $\Gamma(-7/6+2)$, as defined by Eq.~\eqref{Euler}. The second equality in Eq.~\eqref{completegamma} follows from applying the recursion relation in Eq.~\eqref{rr} $two$ times to $\Gamma(-7/6+2)$. In summary, to obtain $I_{3} = \Gamma(-7/6)$ for $N=1$, we performed \mbox{$N+1 = 2$} integrations by parts on $I_{3}$ in order to guarantee that $I_{3}$ could be expressed in terms of a complete gamma function with positive argument. To this function, we then applied Eq.~\eqref{rr} $N+1 = 2$ times to achieve $\Gamma(-7/6)$, which shows that Eq.~\eqref{I123} is identical in form to  Eq.~\eqref{seriesapp} for $p=-7/6$ as long as $\Gamma(-7/6)$ in Eq.~\eqref{I123} is given by Eq.~\eqref{rr}. Therefore, we can apply Eq.~\eqref{seriesapp} for $\Gamma(-7/6,s)$ which gives Eq.~\eqref{series} for $1-\lambda = -7/6$,
\begin{equation*}
\Gamma (-7/6, s) = \Gamma(-7/6) - \frac{s^{-7/6}}{(-7/6)}\,+{\cal O}(s^{-1/6}).
\end{equation*}

The iterative procedure outlined above for $p= -7/6$ and $N=1$ generalizes to arbitrary noninteger values of \mbox{$p<0$}.  No matter how negative such a value of $p$ is, integrating $I_{3}$ by parts $N+1$ times generates a \mbox{$\Gamma(p+N+1)$}, where $(p+N+1) > 0$, and the boundary term always vanishes at $t=\infty$ and $t=0$ for reasons similar to those stated above: the $t^{-7/6} (1-t)$ term in Eq.~\eqref{vanishatinfinity} generalizes to a contribution that is proportional to $\sum_{0}^{N} t^{p+n}$, which vanishes at $t=\infty$ since $(p+n)$ remains negative for all $0 \le n \le N$, and the sum in Eq.~\eqref{boundaryterm} vanishes at $t=0$ because it attains a new lower limit such that the smallest power of $t$ in Eq.~\eqref{boundaryterm} becomes $(p+N+1)$, which is positive. Then, Eq.~\eqref{rr} can be applied to $\Gamma(p+N+1)$ $N+1$ times to produce $I_{3} = \Gamma(p)$. Therefore, Eq.~\eqref{seriesapp} provides a series expansion for $\Gamma(p,s)$ if $p<0$ and noninteger, provided that $\Gamma(p)$ is defined by Eq.~\eqref{rr}.


\begin{thebibliography}{42}
\expandafter\ifx\csname natexlab\endcsname\relax\def\natexlab#1{#1}\fi
\expandafter\ifx\csname bibnamefont\endcsname\relax
  \def\bibnamefont#1{#1}\fi
\expandafter\ifx\csname bibfnamefont\endcsname\relax
  \def\bibfnamefont#1{#1}\fi
\expandafter\ifx\csname citenamefont\endcsname\relax
  \def\citenamefont#1{#1}\fi
\expandafter\ifx\csname url\endcsname\relax
  \def\url#1{\texttt{#1}}\fi
\expandafter\ifx\csname urlprefix\endcsname\relax\def\urlprefix{URL }\fi
\providecommand{\bibinfo}[2]{#2}
\providecommand{\eprint}[2][]{\url{#2}}

\bibitem[{\citenamefont{Hofmann et~al.}(2001)\citenamefont{Hofmann, Schwarz,
  and Stoecker}}]{Hofmann2001}
\bibinfo{author}{\bibfnamefont{S.}~\bibnamefont{Hofmann}},
  \bibinfo{author}{\bibfnamefont{D.~J.} \bibnamefont{Schwarz}},
  \bibnamefont{and} \bibinfo{author}{\bibfnamefont{H.}~\bibnamefont{Stoecker}},
  \bibinfo{journal}{Phys. Rev.} \textbf{\bibinfo{volume}{D64}},
  \bibinfo{pages}{083507} (\bibinfo{year}{2001}), \eprint{astro-ph/0104173}.

\bibitem[{\citenamefont{Bringmann and Hofmann}(2007)}]{Bringmann2007}
\bibinfo{author}{\bibfnamefont{T.}~\bibnamefont{Bringmann}} \bibnamefont{and}
  \bibinfo{author}{\bibfnamefont{S.}~\bibnamefont{Hofmann}},
  \bibinfo{journal}{JCAP} \textbf{\bibinfo{volume}{0704}}, \bibinfo{pages}{016}
  (\bibinfo{year}{2007}), \bibinfo{note}{[Erratum: JCAP1603,no.03,E02(2016)]},
  \eprint{hep-ph/0612238}.

\bibitem[{\citenamefont{Bringmann}(2009)}]{Bringmann2009}
\bibinfo{author}{\bibfnamefont{T.}~\bibnamefont{Bringmann}},
  \bibinfo{journal}{New J. Phys.} \textbf{\bibinfo{volume}{11}},
  \bibinfo{pages}{105027} (\bibinfo{year}{2009}), \eprint{0903.0189}.

\bibitem[{\citenamefont{Visinelli and Gondolo}(2015)}]{Visinelli2015}
\bibinfo{author}{\bibfnamefont{L.}~\bibnamefont{Visinelli}} \bibnamefont{and}
  \bibinfo{author}{\bibfnamefont{P.}~\bibnamefont{Gondolo}},
  \bibinfo{journal}{Phys. Rev.} \textbf{\bibinfo{volume}{D91}},
  \bibinfo{pages}{083526} (\bibinfo{year}{2015}), \eprint{1501.02233}.

\bibitem[{\citenamefont{Waldstein and Erickcek}(2017)}]{WE2017}
\bibinfo{author}{\bibfnamefont{I.~R.} \bibnamefont{Waldstein}}
  \bibnamefont{and} \bibinfo{author}{\bibfnamefont{A.~L.}
  \bibnamefont{Erickcek}}, \bibinfo{journal}{Phys. Rev. D}
  \textbf{\bibinfo{volume}{95}}, \bibinfo{pages}{088301}
  (\bibinfo{year}{2017}),
  \urlprefix\url{https://link.aps.org/doi/10.1103/PhysRevD.95.088301}.

\bibitem[{\citenamefont{Green et~al.}(2004)\citenamefont{Green, Hofmann, and
  Schwarz}}]{Green2004}
\bibinfo{author}{\bibfnamefont{A.~M.} \bibnamefont{Green}},
  \bibinfo{author}{\bibfnamefont{S.}~\bibnamefont{Hofmann}}, \bibnamefont{and}
  \bibinfo{author}{\bibfnamefont{D.~J.} \bibnamefont{Schwarz}},
  \bibinfo{journal}{Mon. Not. Roy. Astron. Soc.}
  \textbf{\bibinfo{volume}{353}}, \bibinfo{pages}{L23} (\bibinfo{year}{2004}),
  \eprint{astro-ph/0309621}.

\bibitem[{\citenamefont{Green et~al.}(2005)\citenamefont{Green, Hofmann, and
  Schwarz}}]{Green2005}
\bibinfo{author}{\bibfnamefont{A.~M.} \bibnamefont{Green}},
  \bibinfo{author}{\bibfnamefont{S.}~\bibnamefont{Hofmann}}, \bibnamefont{and}
  \bibinfo{author}{\bibfnamefont{D.~J.} \bibnamefont{Schwarz}},
  \bibinfo{journal}{JCAP} \textbf{\bibinfo{volume}{0508}}, \bibinfo{pages}{003}
  (\bibinfo{year}{2005}), \eprint{astro-ph/0503387}.

\bibitem[{\citenamefont{Loeb and Zaldarriaga}(2005)}]{Loeb2005}
\bibinfo{author}{\bibfnamefont{A.}~\bibnamefont{Loeb}} \bibnamefont{and}
  \bibinfo{author}{\bibfnamefont{M.}~\bibnamefont{Zaldarriaga}},
  \bibinfo{journal}{Phys. Rev.} \textbf{\bibinfo{volume}{D71}},
  \bibinfo{pages}{103520} (\bibinfo{year}{2005}), \eprint{astro-ph/0504112}.

\bibitem[{\citenamefont{Bertschinger}(2006)}]{Bert2006}
\bibinfo{author}{\bibfnamefont{E.}~\bibnamefont{Bertschinger}},
  \bibinfo{journal}{Phys. Rev.} \textbf{\bibinfo{volume}{D74}},
  \bibinfo{pages}{063509} (\bibinfo{year}{2006}), \eprint{astro-ph/0607319}.

\bibitem[{\citenamefont{Gondolo et~al.}(2012)\citenamefont{Gondolo, Hisano, and
  Kadota}}]{Gondolo2012}
\bibinfo{author}{\bibfnamefont{P.}~\bibnamefont{Gondolo}},
  \bibinfo{author}{\bibfnamefont{J.}~\bibnamefont{Hisano}}, \bibnamefont{and}
  \bibinfo{author}{\bibfnamefont{K.}~\bibnamefont{Kadota}},
  \bibinfo{journal}{Phys. Rev.} \textbf{\bibinfo{volume}{D86}},
  \bibinfo{pages}{083523} (\bibinfo{year}{2012}), \eprint{1205.1914}.

\bibitem[{\citenamefont{Profumo et~al.}(2006)\citenamefont{Profumo, Sigurdson,
  and Kamionkowski}}]{Profumo2006}
\bibinfo{author}{\bibfnamefont{S.}~\bibnamefont{Profumo}},
  \bibinfo{author}{\bibfnamefont{K.}~\bibnamefont{Sigurdson}},
  \bibnamefont{and}
  \bibinfo{author}{\bibfnamefont{M.}~\bibnamefont{Kamionkowski}},
  \bibinfo{journal}{Phys. Rev. Lett.} \textbf{\bibinfo{volume}{97}},
  \bibinfo{pages}{031301} (\bibinfo{year}{2006}), \eprint{astro-ph/0603373}.

\bibitem[{\citenamefont{Gelmini and Gondolo}(2008)}]{Gelmini2008}
\bibinfo{author}{\bibfnamefont{G.~B.} \bibnamefont{Gelmini}} \bibnamefont{and}
  \bibinfo{author}{\bibfnamefont{P.}~\bibnamefont{Gondolo}},
  \bibinfo{journal}{JCAP} \textbf{\bibinfo{volume}{0810}}, \bibinfo{pages}{002}
  (\bibinfo{year}{2008}), \eprint{0803.2349}.

\bibitem[{\citenamefont{Arcadi and Ullio}(2011)}]{Arcadi2011}
\bibinfo{author}{\bibfnamefont{G.}~\bibnamefont{Arcadi}} \bibnamefont{and}
  \bibinfo{author}{\bibfnamefont{P.}~\bibnamefont{Ullio}},
  \bibinfo{journal}{Phys. Rev.} \textbf{\bibinfo{volume}{D84}},
  \bibinfo{pages}{043520} (\bibinfo{year}{2011}), \eprint{1104.3591}.

\bibitem[{\citenamefont{Kane et~al.}(2016)\citenamefont{Kane, Kumar, Nelson,
  and Zheng}}]{Kane22015}
\bibinfo{author}{\bibfnamefont{G.~L.} \bibnamefont{Kane}},
  \bibinfo{author}{\bibfnamefont{P.}~\bibnamefont{Kumar}},
  \bibinfo{author}{\bibfnamefont{B.~D.} \bibnamefont{Nelson}},
  \bibnamefont{and} \bibinfo{author}{\bibfnamefont{B.}~\bibnamefont{Zheng}},
  \bibinfo{journal}{Phys. Rev.} \textbf{\bibinfo{volume}{D93}},
  \bibinfo{pages}{063527} (\bibinfo{year}{2016}), \eprint{1502.05406}.

\bibitem[{\citenamefont{Erickcek and Sigurdson}(2011)}]{Erickcek2011}
\bibinfo{author}{\bibfnamefont{A.~L.} \bibnamefont{Erickcek}} \bibnamefont{and}
  \bibinfo{author}{\bibfnamefont{K.}~\bibnamefont{Sigurdson}},
  \bibinfo{journal}{Phys. Rev.} \textbf{\bibinfo{volume}{D84}},
  \bibinfo{pages}{083503} (\bibinfo{year}{2011}), \eprint{1106.0536}.

\bibitem[{\citenamefont{Barenboim and Rasero}(2014)}]{Barenboim2014}
\bibinfo{author}{\bibfnamefont{G.}~\bibnamefont{Barenboim}} \bibnamefont{and}
  \bibinfo{author}{\bibfnamefont{J.}~\bibnamefont{Rasero}},
  \bibinfo{journal}{JHEP} \textbf{\bibinfo{volume}{04}}, \bibinfo{pages}{138}
  (\bibinfo{year}{2014}), \eprint{1311.4034}.

\bibitem[{\citenamefont{{Fan} et~al.}(2014)\citenamefont{{Fan}, {{\"O}zsoy},
  and {Watson}}}]{Fan2014}
\bibinfo{author}{\bibfnamefont{J.}~\bibnamefont{{Fan}}},
  \bibinfo{author}{\bibfnamefont{O.}~\bibnamefont{{{\"O}zsoy}}},
  \bibnamefont{and} \bibinfo{author}{\bibfnamefont{S.}~\bibnamefont{{Watson}}},
  \bibinfo{journal}{\prd} \textbf{\bibinfo{volume}{90}}, \bibinfo{eid}{043536}
  (\bibinfo{year}{2014}), \eprint{1405.7373}.

\bibitem[{\citenamefont{Erickcek}(2015)}]{Erickcek2015long}
\bibinfo{author}{\bibfnamefont{A.~L.} \bibnamefont{Erickcek}},
  \bibinfo{journal}{Phys. Rev.} \textbf{\bibinfo{volume}{D92}},
  \bibinfo{pages}{103505} (\bibinfo{year}{2015}), \eprint{1504.03335}.

\bibitem[{\citenamefont{Erickcek et~al.}(2015)\citenamefont{Erickcek, Sinha,
  and Watson}}]{Erickcek2015short}
\bibinfo{author}{\bibfnamefont{A.~L.} \bibnamefont{Erickcek}},
  \bibinfo{author}{\bibfnamefont{K.}~\bibnamefont{Sinha}}, \bibnamefont{and}
  \bibinfo{author}{\bibfnamefont{S.}~\bibnamefont{Watson}}
  (\bibinfo{year}{2015}), \eprint{1510.04291}.

\bibitem[{\citenamefont{Kawasaki et~al.}(1999)\citenamefont{Kawasaki, Kohri,
  and Sugiyama}}]{Kawasaki1999}
\bibinfo{author}{\bibfnamefont{M.}~\bibnamefont{Kawasaki}},
  \bibinfo{author}{\bibfnamefont{K.}~\bibnamefont{Kohri}}, \bibnamefont{and}
  \bibinfo{author}{\bibfnamefont{N.}~\bibnamefont{Sugiyama}},
  \bibinfo{journal}{Phys. Rev. Lett.} \textbf{\bibinfo{volume}{82}},
  \bibinfo{pages}{4168} (\bibinfo{year}{1999}), \eprint{astro-ph/9811437}.

\bibitem[{\citenamefont{Kawasaki et~al.}(2000)\citenamefont{Kawasaki, Kohri,
  and Sugiyama}}]{Kawasaki2000}
\bibinfo{author}{\bibfnamefont{M.}~\bibnamefont{Kawasaki}},
  \bibinfo{author}{\bibfnamefont{K.}~\bibnamefont{Kohri}}, \bibnamefont{and}
  \bibinfo{author}{\bibfnamefont{N.}~\bibnamefont{Sugiyama}},
  \bibinfo{journal}{Phys. Rev.} \textbf{\bibinfo{volume}{D62}},
  \bibinfo{pages}{023506} (\bibinfo{year}{2000}), \eprint{astro-ph/0002127}.

\bibitem[{\citenamefont{Hannestad}(2004)}]{Hannestad2004}
\bibinfo{author}{\bibfnamefont{S.}~\bibnamefont{Hannestad}},
  \bibinfo{journal}{Phys. Rev.} \textbf{\bibinfo{volume}{D70}},
  \bibinfo{pages}{043506} (\bibinfo{year}{2004}), \eprint{astro-ph/0403291}.

\bibitem[{\citenamefont{Ichikawa et~al.}(2005)\citenamefont{Ichikawa, Kawasaki,
  and Takahashi}}]{Ichikawa2005}
\bibinfo{author}{\bibfnamefont{K.}~\bibnamefont{Ichikawa}},
  \bibinfo{author}{\bibfnamefont{M.}~\bibnamefont{Kawasaki}}, \bibnamefont{and}
  \bibinfo{author}{\bibfnamefont{F.}~\bibnamefont{Takahashi}},
  \bibinfo{journal}{Phys. Rev.} \textbf{\bibinfo{volume}{D72}},
  \bibinfo{pages}{043522} (\bibinfo{year}{2005}), \eprint{astro-ph/0505395}.

\bibitem[{\citenamefont{Coughlan et~al.}(1983)\citenamefont{Coughlan, Fischler,
  Kolb, Raby, and Ross}}]{Coughlan1983}
\bibinfo{author}{\bibfnamefont{G.~D.} \bibnamefont{Coughlan}},
  \bibinfo{author}{\bibfnamefont{W.}~\bibnamefont{Fischler}},
  \bibinfo{author}{\bibfnamefont{E.~W.} \bibnamefont{Kolb}},
  \bibinfo{author}{\bibfnamefont{S.}~\bibnamefont{Raby}}, \bibnamefont{and}
  \bibinfo{author}{\bibfnamefont{G.~G.} \bibnamefont{Ross}},
  \bibinfo{journal}{Phys. Lett.} \textbf{\bibinfo{volume}{B131}},
  \bibinfo{pages}{59} (\bibinfo{year}{1983}).

\bibitem[{\citenamefont{de~Carlos et~al.}(1993)\citenamefont{de~Carlos, Casas,
  Quevedo, and Roulet}}]{deCarlos1984}
\bibinfo{author}{\bibfnamefont{B.}~\bibnamefont{de~Carlos}},
  \bibinfo{author}{\bibfnamefont{J.~A.} \bibnamefont{Casas}},
  \bibinfo{author}{\bibfnamefont{F.}~\bibnamefont{Quevedo}}, \bibnamefont{and}
  \bibinfo{author}{\bibfnamefont{E.}~\bibnamefont{Roulet}},
  \bibinfo{journal}{Phys. Lett.} \textbf{\bibinfo{volume}{B318}},
  \bibinfo{pages}{447} (\bibinfo{year}{1993}), \eprint{hep-ph/9308325}.

\bibitem[{\citenamefont{Banks et~al.}(1994)\citenamefont{Banks, Kaplan, and
  Nelson}}]{Banks1994}
\bibinfo{author}{\bibfnamefont{T.}~\bibnamefont{Banks}},
  \bibinfo{author}{\bibfnamefont{D.~B.} \bibnamefont{Kaplan}},
  \bibnamefont{and} \bibinfo{author}{\bibfnamefont{A.~E.}
  \bibnamefont{Nelson}}, \bibinfo{journal}{Phys. Rev.}
  \textbf{\bibinfo{volume}{D49}}, \bibinfo{pages}{779} (\bibinfo{year}{1994}),
  \eprint{hep-ph/9308292}.

\bibitem[{\citenamefont{Banks et~al.}(1995{\natexlab{a}})\citenamefont{Banks,
  Berkooz, and Steinhardt}}]{Banks1995_1}
\bibinfo{author}{\bibfnamefont{T.}~\bibnamefont{Banks}},
  \bibinfo{author}{\bibfnamefont{M.}~\bibnamefont{Berkooz}}, \bibnamefont{and}
  \bibinfo{author}{\bibfnamefont{P.~J.} \bibnamefont{Steinhardt}},
  \bibinfo{journal}{Phys. Rev.} \textbf{\bibinfo{volume}{D52}},
  \bibinfo{pages}{705} (\bibinfo{year}{1995}{\natexlab{a}}),
  \eprint{hep-th/9501053}.

\bibitem[{\citenamefont{Banks et~al.}(1995{\natexlab{b}})\citenamefont{Banks,
  Berkooz, Shenker, Moore, and Steinhardt}}]{Banks1995_2}
\bibinfo{author}{\bibfnamefont{T.}~\bibnamefont{Banks}},
  \bibinfo{author}{\bibfnamefont{M.}~\bibnamefont{Berkooz}},
  \bibinfo{author}{\bibfnamefont{S.~H.} \bibnamefont{Shenker}},
  \bibinfo{author}{\bibfnamefont{G.~W.} \bibnamefont{Moore}}, \bibnamefont{and}
  \bibinfo{author}{\bibfnamefont{P.~J.} \bibnamefont{Steinhardt}},
  \bibinfo{journal}{Phys. Rev.} \textbf{\bibinfo{volume}{D52}},
  \bibinfo{pages}{3548} (\bibinfo{year}{1995}{\natexlab{b}}),
  \eprint{hep-th/9503114}.

\bibitem[{\citenamefont{Acharya et~al.}(2008)\citenamefont{Acharya, Kumar,
  Bobkov, Kane, Shao, and Watson}}]{Acharya2008}
\bibinfo{author}{\bibfnamefont{B.~S.} \bibnamefont{Acharya}},
  \bibinfo{author}{\bibfnamefont{P.}~\bibnamefont{Kumar}},
  \bibinfo{author}{\bibfnamefont{K.}~\bibnamefont{Bobkov}},
  \bibinfo{author}{\bibfnamefont{G.}~\bibnamefont{Kane}},
  \bibinfo{author}{\bibfnamefont{J.}~\bibnamefont{Shao}}, \bibnamefont{and}
  \bibinfo{author}{\bibfnamefont{S.}~\bibnamefont{Watson}},
  \bibinfo{journal}{JHEP} \textbf{\bibinfo{volume}{06}}, \bibinfo{pages}{064}
  (\bibinfo{year}{2008}), \eprint{0804.0863}.

\bibitem[{\citenamefont{Acharya et~al.}(2009)\citenamefont{Acharya, Kane,
  Watson, and Kumar}}]{Acharya2009}
\bibinfo{author}{\bibfnamefont{B.~S.} \bibnamefont{Acharya}},
  \bibinfo{author}{\bibfnamefont{G.}~\bibnamefont{Kane}},
  \bibinfo{author}{\bibfnamefont{S.}~\bibnamefont{Watson}}, \bibnamefont{and}
  \bibinfo{author}{\bibfnamefont{P.}~\bibnamefont{Kumar}},
  \bibinfo{journal}{Phys. Rev.} \textbf{\bibinfo{volume}{D80}},
  \bibinfo{pages}{083529} (\bibinfo{year}{2009}), \eprint{0908.2430}.

\bibitem[{\citenamefont{Blinov et~al.}(2015)\citenamefont{Blinov, Kozaczuk,
  Menon, and Morrissey}}]{Blinov2014}
\bibinfo{author}{\bibfnamefont{N.}~\bibnamefont{Blinov}},
  \bibinfo{author}{\bibfnamefont{J.}~\bibnamefont{Kozaczuk}},
  \bibinfo{author}{\bibfnamefont{A.}~\bibnamefont{Menon}}, \bibnamefont{and}
  \bibinfo{author}{\bibfnamefont{D.~E.} \bibnamefont{Morrissey}},
  \bibinfo{journal}{Phys. Rev.} \textbf{\bibinfo{volume}{D91}},
  \bibinfo{pages}{035026} (\bibinfo{year}{2015}), \eprint{1409.1222}.

\bibitem[{\citenamefont{Kane et~al.}(2015)\citenamefont{Kane, Sinha, and
  Watson}}]{Kane2015}
\bibinfo{author}{\bibfnamefont{G.}~\bibnamefont{Kane}},
  \bibinfo{author}{\bibfnamefont{K.}~\bibnamefont{Sinha}}, \bibnamefont{and}
  \bibinfo{author}{\bibfnamefont{S.}~\bibnamefont{Watson}},
  \bibinfo{journal}{Int. J. Mod. Phys.} \textbf{\bibinfo{volume}{D24}},
  \bibinfo{pages}{1530022} (\bibinfo{year}{2015}), \eprint{1502.07746}.

\bibitem[{\citenamefont{Tenkanen and Vaskonen}(2016)}]{Tenkanen2016}
\bibinfo{author}{\bibfnamefont{T.}~\bibnamefont{Tenkanen}} \bibnamefont{and}
  \bibinfo{author}{\bibfnamefont{V.}~\bibnamefont{Vaskonen}},
  \bibinfo{journal}{Phys. Rev.} \textbf{\bibinfo{volume}{D94}},
  \bibinfo{pages}{083516} (\bibinfo{year}{2016}), \eprint{1606.00192}.

\bibitem[{\citenamefont{Zhang}(2015)}]{Zhang2015}
\bibinfo{author}{\bibfnamefont{Y.}~\bibnamefont{Zhang}},
  \bibinfo{journal}{JCAP} \textbf{\bibinfo{volume}{1505}}, \bibinfo{pages}{008}
  (\bibinfo{year}{2015}), \eprint{1502.06983}.

\bibitem[{\citenamefont{Choi et~al.}(2015)\citenamefont{Choi, Gong, and
  Shin}}]{Choi2015}
\bibinfo{author}{\bibfnamefont{K.-Y.} \bibnamefont{Choi}},
  \bibinfo{author}{\bibfnamefont{J.-O.} \bibnamefont{Gong}}, \bibnamefont{and}
  \bibinfo{author}{\bibfnamefont{C.~S.} \bibnamefont{Shin}},
  \bibinfo{journal}{Phys. Rev. Lett.} \textbf{\bibinfo{volume}{115}},
  \bibinfo{pages}{211302} (\bibinfo{year}{2015}), \eprint{1507.03871}.

\bibitem[{\citenamefont{Georg et~al.}(2016)\citenamefont{Georg, {\c
  S}eng{\"o}r, and Watson}}]{Georg2016}
\bibinfo{author}{\bibfnamefont{J.}~\bibnamefont{Georg}},
  \bibinfo{author}{\bibfnamefont{G.}~\bibnamefont{{\c S}eng{\"o}r}},
  \bibnamefont{and} \bibinfo{author}{\bibfnamefont{S.}~\bibnamefont{Watson}},
  \bibinfo{journal}{Phys. Rev.} \textbf{\bibinfo{volume}{D93}},
  \bibinfo{pages}{123523} (\bibinfo{year}{2016}), \eprint{1603.00023}.

\bibitem[{\citenamefont{Giudice et~al.}(2001)\citenamefont{Giudice, Kolb, and
  Riotto}}]{Giudice2001}
\bibinfo{author}{\bibfnamefont{G.~F.} \bibnamefont{Giudice}},
  \bibinfo{author}{\bibfnamefont{E.~W.} \bibnamefont{Kolb}}, \bibnamefont{and}
  \bibinfo{author}{\bibfnamefont{A.}~\bibnamefont{Riotto}},
  \bibinfo{journal}{Phys. Rev.} \textbf{\bibinfo{volume}{D64}},
  \bibinfo{pages}{023508} (\bibinfo{year}{2001}), \eprint{hep-ph/0005123}.

\bibitem[{\citenamefont{D'Eramo et~al.}(2017)\citenamefont{D'Eramo, Fernandez,
  and Profumo}}]{Profumo2017}
\bibinfo{author}{\bibfnamefont{F.}~\bibnamefont{D'Eramo}},
  \bibinfo{author}{\bibfnamefont{N.}~\bibnamefont{Fernandez}},
  \bibnamefont{and} \bibinfo{author}{\bibfnamefont{S.}~\bibnamefont{Profumo}}
  (\bibinfo{year}{2017}), \eprint{1703.04793}.

\bibitem[{\citenamefont{Armendariz-Picon
  et~al.}(1999)\citenamefont{Armendariz-Picon, Damour, and
  Mukhanov}}]{Armendariz1999}
\bibinfo{author}{\bibfnamefont{C.}~\bibnamefont{Armendariz-Picon}},
  \bibinfo{author}{\bibfnamefont{T.}~\bibnamefont{Damour}}, \bibnamefont{and}
  \bibinfo{author}{\bibfnamefont{V.~F.} \bibnamefont{Mukhanov}},
  \bibinfo{journal}{Phys. Lett.} \textbf{\bibinfo{volume}{B458}},
  \bibinfo{pages}{209} (\bibinfo{year}{1999}), \eprint{hep-th/9904075}.

\bibitem[{\citenamefont{Christopherson and Malik}(2009)}]{Christopherson2008}
\bibinfo{author}{\bibfnamefont{A.~J.} \bibnamefont{Christopherson}}
  \bibnamefont{and} \bibinfo{author}{\bibfnamefont{K.~A.} \bibnamefont{Malik}},
  \bibinfo{journal}{Phys. Lett.} \textbf{\bibinfo{volume}{B675}},
  \bibinfo{pages}{159} (\bibinfo{year}{2009}), \eprint{0809.3518}.

\bibitem[{\citenamefont{Waldstein et~al.}(2017)\citenamefont{Waldstein, Ilie,
  and Erickcek}}]{WIE2016}
\bibinfo{author}{\bibfnamefont{I.~R.} \bibnamefont{Waldstein}},
  \bibinfo{author}{\bibfnamefont{C.}~\bibnamefont{Ilie}}, \bibnamefont{and}
  \bibinfo{author}{\bibfnamefont{A.~L.} \bibnamefont{Erickcek}},
  \bibinfo{journal}{in preparation}  \bibinfo{year}{(2017).}

\bibitem[{\citenamefont{Bender and Orszag}(1999)}]{Bender1978}
\bibinfo{author}{\bibfnamefont{C.~M.} \bibnamefont{Bender}} \bibnamefont{and}
  \bibinfo{author}{\bibfnamefont{S.~A.} \bibnamefont{Orszag}},
  \emph{\bibinfo{title}{Advanced Mathematical Methods for Scientists and
  Engineers Asymptotic Methods and Perturbation Theory}}
  (\bibinfo{publisher}{Springer}, \bibinfo{year}{1999}).

\end{thebibliography}
\end{document}